# Unraveling ferroelectric polarization and ionic contributions to electroresistance in epitaxial $Hf_{0.5}Zr_{0.5}O_2$ tunnel junctions

Milena Cervo Sulzbach[1], Saúl Estandía[1], Xiao Long[1], Jike Lyu[1], Nico Dix[1], Jaume Gàzquez[1], Matthew F. Chisholm[2], Florencio Sánchez[1], Ignasi Fina[1]*, Josep Fontcuberta[1]*

[1] Institut de Ciència de Materials de Barcelona (ICMAB-CSIC), Campus UAB, Bellaterra, Catalonia 08193, Spain

[2] Center for Nanophase Materials Sciences, Oak Ridge National Laboratory, Tennessee 37831-6064, USA

E-mail: ifina@icmab.es, fontcuberta@icmab.cat

Tunnel devices based on ferroelectric $Hf_{0.5}Zr_{0.5}O_2$ (HZO) barriers hold great promises for emerging data storage and computing technologies. The resistance state of the device can be changed by a suitable writing voltage. However, the microscopic mechanisms leading to the resistance change are an intricate interplay between ferroelectric polarization controlled barrier properties and defect-related transport mechanisms. Here is shown the fundamental role of the microstructure of HZO films setting the balance between those contributions. The oxide film presents coherent or incoherent grain boundaries, associated to the existence of monoclinic and orthorhombic phases in HZO films, which are dictated by the mismatch with the substrates for epitaxial growth. These grain boundaries are the toggle that allows to obtain either large (up to $\approx$ 450 %) and fully reversible genuine polarization controlled electroresistance when only the orthorhombic phase is present or an irreversible and extremely large ($\approx 10^3$-$10^5$ %) electroresistance when both phases coexist.

**Introduction**

Ferroelectric tunnel devices are driving attention as class of resistive switching devices due to its low energy consumption, fast writing/reading and small cell size.[1,2] A ferroelectric tunnel junction (FeTJ) has a capacitor-like structure in which a nanometric ferroelectric layer is sandwiched between two metallic electrodes. The height of the energy barrier ($\Phi$) at the ferroelectric/electrode interfaces depends on the direction of the ferroelectric polarization ($P$) and the difference of electrodes (i and j) screening lengths ($\delta_{i,j}$). For a given pair of electrodes, having $\delta_i \neq \delta_j$, $\Phi(P(\uparrow))$ and $\Phi(P(\downarrow))$ are different, therefore two distinct resistance states can be obtained for the device when polarizing the ferroelectric barrier in opposite directions.[3] The change in resistance, so-called electroresistance ($ER$), is quantified by $ER = [R(V_W^+)-R(V_W^-)]/R_{min}(V_W^{+,-})$, where $R(V_W^+)$ and $R(V_W^-)$ are the resistances after polarizing the junction with writing voltages $V_W^+$ or $V_W^-$ and $R_{min}(V_W^{+,-})$ is the minimum resistance among these states. Accordingly, binary high (OFF) and low resistance (ON) states can be written in a ferroelectric memory cell and read by probing its resistance. It has also been shown that by performing minor polarization loops, ferroelectric tunnel devices can store information in different resistive states, mimicking the functioning of a memristive element.[4,5] This approach has been successfully achieved by using ferroelectric perovskites such as $BaTiO_3$,[6–9] $Pb(Zr_{0.2}Ti_{0.8})O_3$[10,11] and $BiFeO_3$.[12–14] If a defect-related depletion layer exists at the ferroelectric/electrode interface, the effective barrier width ($W$) will be modified when reversing the ferroelectric polarization direction. Therefore it will provide an additional contribution to the $ER$, as reported in $BaTiO_3$ tunnel barriers using $La_{2/3}Sr_{1/3}MnO_3$ (LSMO) or n-Si as electrodes.[15,16]

The discovery of ferroelectricity in $HfO_2$-based thin films,[17] being CMOS-compatible, has created expectations that new functionalities could be implemented in current silicon-based platforms.[18,19] Epitaxial growth of ferroelectric $Hf_{0.5}Zr_{0.5}O_2$ (HZO) thin films has been achieved,[20–26] opening the door to engineer epitaxial FeTJ and magnetic-FeTJ.[25] The ferroelectric character of ultrathin (few nanometer layers) epitaxial HZO films has been assessed by piezoelectric force microscopy and by recording P-E loops on micrometric electrodes.[20–26] Thus, a natural question arises: can ferroelectric-related $ER$ be observed in epitaxial $Hf_{0.5}Zr_{0.5}O_2$ tunnel barriers and exploited to create $HfO_2$-based FeTJ?

A key point is that the $ER$ of the ferroelectric tunnel junctions is not only sensitive to polarization effects described above, but may also be affected by ionic or other defect-related charge motion at the barrier. In the most studied ferroelectric tunnel barriers involving $BaTiO_3$ and $BiFeO_3$, for instance, it has been recognized that ionic motion plays a role in $ER$.[4,27–29] Indeed, electroresistance in non-ferroelectric oxides, including $HfO_2$, has been largely explored to create resistive switching[30] memories and logic elements,[31,32] taking advantage of the high (OFF) and low (ON) resistance states that can be obtained by applying suitable voltage pulses.[32,33] It has been shown that after a voltage-induced *forming* step, reversible OFF and ON states can be obtained (*set* and *reset*) by cycling the external voltage. The microscopic mechanisms for *forming* and *set-reset* processes have been much discussed. For example, Bersuker et al[34] showed that, in monoclinic $HfO_2$, the $ER$ can be modeled by a *forming* step related to the dielectric soft breakdown in a filamentary shape. In the formation of conductive filaments, the $HfO_2$ grain boundaries show a crucial role. The filaments produced by the soft breakdown are formed at $HfO_2$ grain boundaries, which involves oxygen dissociation (oxygen vacancy generation) and the creation of a conducting channel.[35] Electric-field driven oxygen vacancies accumulation at the electrode forms an conducting layer which corresponds to the ON state. When the electric field switches the polarity, the oxygen vacancies diffuse through the oxide, restoring the insulating character of the interfacial layer and ultimately producing the OFF state.[34]

Recently, $ER$ in tunnel junctions involving epitaxial ferroelectric $HfO_2$ barriers have been reported.[25,26,36,37] However, available data do not allow indisputably to conclude if the observed *ER* is directly governed by polarization-related effects or it results from electric-field induced charge motion, as filamentary conductivity or other mechanism, as commonly observed in non-tunneling HfO₂ barriers.[38] For instance, Wei et al. reported current-voltage curves (I-V) of 2 nm HZO barriers and argued that data can be described using the Brinkman model,[39] indicating direct tunnel transport across the barrier.[25] However, the dependence of the barrier properties on the polarization direction was not analyzed, thus the connection between the measured *ER* and *P* remained undisclosed. Ambriz-Vargas et al.[36,37] reported also I-V measurements on HZO (2.8 nm thick) barriers. It was shown that the I-V curves are dramatically different after a writing voltage of ± 2.2 V. Nonetheless, the data were analyzed in a narrow voltage region, in which the I-V curves were almost linear, and precluding robust extraction of barrier parameters and its possible change due to polarization reversal. Yoong et al.[26] also reported I-V data on HZO epitaxial tunnel barriers about 10 nm thick and their polarization *P*(*V*) loops. Transport data were analyzed in term of thermionic emission over a polarization-modulated barrier. The use of a thermionic model was adequate because direct tunneling across such thick barrier is prohibited. It was observed that the properties of the barriers were significantly different depending on the magnitude of the writing voltage, whereas, in the same voltage range, the polarization loop appeared to be saturated. This observation suggests that barrier properties are not only dictated by polarization but other mechanism may come at play.

Here we address this crucial issue by recording the polarization *P*(*V*) and *ER*(*V*$_W$) loops on epitaxial ferroelectric HZO tunnel barriers grown on single crystalline cubic substrates with different structural mismatch with HZO. The rationale behind is the following. First, we take advantage of the recent discovery by Estandía et al.[40] that the structural mismatch between HZO and LSMO-coated substrates determines the microstructure of the HZO films. This structure must be interpreted as the relative abundance of orthorhombic (o-HZO) (ferroelectric) and monoclinic (m-HZO) (non-ferroelectric) phases and the fine details of the unavoidable grain boundaries. Second, incoherent grain boundaries in polycrystalline $HfO_2$ are known to be instrumental for electric field controlled resistive switching.[34] Therefore, varying the relative amount of coherent and non-coherent grain boundaries, the latter arises from the coexistence of o- and m-HZO phases and their respective role on electroresistance can be disentangled. It will be shown that HZO films grown on $GdScO_3$ and $TbScO_3$ display genuine polarization-related electroresistance (up to $\approx 450$ %) developing at the coercive voltage of the film. Whereas HZO films on $SrTiO_3$ display a coexistence of ferroelectric and non-ferroelectric resistive switching channels, operating at different voltage ranges and with electroresistance of up to 450 % and $10^4$-$10^5$ %, respectively. Therefore, ferroelectric polarization and defect-related effects contribute to the *ER* in HZO ultrathin ferroelectric films. The structural mismatch between HZO and substrate and the resulting HZO microstructure are the toggle that sets the balance between polarization-related and defect-related mechanisms governing the device *ER*.

**Results**

Epitaxial HZO films with nominal thicknesses 4.6 nm were grown on LSMO (22 nm thick) conducting buffer (001) single crystalline substrates $SrTiO_3$ (STO), $GdScO_3$ (GSO), $TbScO_3$ (TSO) (pseudocubic setting is used for scandates) and $(LaAlO_3)_{0.3}(Sr_2TaAlO_6)_{0.7}$ (LSAT) by pulsed laser deposition (PLD). Ex-situ sputtered circular Pt top electrodes of 20 μm of diameter were grown through masks. The heterostructure is sketched in Figure 1a. In Figure 1b, we show an X-ray diffraction 2θ-χ frame and the corresponding integrated θ-2θ scan of the sample grown on STO. The intense (00l) reflections of substrate and LSMO can be observed as well as the (111) reflection of the ferroelectric o-HZO, indicating its prevalence in the film. No other reflections can be seen. The I-V curve recorded on a fresh Pt/HZO/LSMO//STO device (junction $J_1$) shown in Figure 1c and the corresponding $P(V)$ loop shown in Figure 1d assess the ferroelectric nature of the film. The coercive voltage extracted from the position of the switching peaks is $V_C^+ \approx V_C^- \approx 2$ V. Data indicates a remnant polarization $P_R \approx 16$ μC cm$^{-2}$. These ferroelectric characteristics are consistent with previous reports.[20–22]

In Figure 2a, we show I-V curves collected at 5 kHz (junction J$_2$). This measurement frequency allows meaningful comparison with the $ER$ measurements recorded using writing pulses of 100 μs. The polarization switching current peak occurs at about 2 V, indicating $V_C^+$ $\approx 2$ V. The current peak observed at around $V_{max}$ in Figure 2a is a spurious effect due to high frequency used in these measurements (Supporting Information S1). The $ER$ measurement (junction J$_3$) sequence is done first applying a writing triangular pulse of amplitude $V_W$ and duration $\tau_W = 100$ μs as indicated in Figure 2b and using the electrical contact configuration shown in Figure 1a. Then, I-V curves are collected, after a delay time $\tau_D = 0.5$ s, by applying a linear $V_R(t)$ pulse in a small voltage range (from -1 V to +1 V). The resistance is determined at $V_R = 0.9$ V. Notice that, to avoid ferroelectric switching during the reading, the maximum reading voltage is smaller than $V_C$ ($\approx 2$ V). The $ER$ data collected cycling $V_W$ up to ±4.5 V is shown in Figure 2c. Data clearly display noticeable changes of resistance developing at about ±2 V, which closely coincide with $V_C^{+,-}$ observed in Figure 2a (as indicated by dashed lines). Moreover, it can be appreciated in Figure 2c that the $R(V_W)$ is reversible (3 consecutive loops are shown) and the $ER$ is 410 % and no forming steps are necessary. In contrast when increasing $V_W$ (± 6 V), the $ER$ becomes larger ($\approx 10^5$ %) but irreversibility emerges (Figure 2d). Therefore, data in Figure 2 clearly indicates that different $ER$ mechanisms occur depending on the writing voltage. The close correspondence of polarization and $ER$ in Figure 2a and 2c strongly suggests that $ER$ is due to polarization-controlled tunnel transport across the HZO barrier if voltages near $V_W \leq \pm 4.5$ V are used. Ferroelectric switching and $ER$ coexists at larger $V_W$, but these are not consequence one from the other, as reported in polycrystalline HfO$_2$ thin films.[38]

In order to get an insight on the origin of reversible and irreversible $ER$ observed, Figure 3a shows the resistance (junction $J_2$) after increasing positive and negative $V_W$ writing steps (pulse sequence is sketched in Supporting Information S2). Stability of resistance after a certain number of pulses with same amplitude are shown in Supporting Information S3. Resistance $R(V_W)$ data in Figure 3a display two clear regions (I and II) separated by a vertical line. In region I, for 2 V $< |V_W| <$ 5 V, the resistance increases with $V_W^+$ ($V_W > 0$) and decreases for $V_W^-$ ($V_W < 0$). Further increase of $V_W$, irrespective of its polarity, produces a large decrease in resistance (region II, Figure 3a) and $ER$ reduction. Between regions I and II, $ER(V_W)$ displays a well pronounced maximum, as shown in Figure 3 in which data collected up to $V_W = 7.5$ V are depicted. The $ER$ values in region I amount up to $\approx$ 340 % ($V_W \approx 4.5$ V) and increase up to 2000 % when entering region II. Note that these $ER$ values are slightly different from those in Figure 2c due to the different pulse train used for the measurements in Figure 2c and 3a (see Supporting Information S2). It is worth mentioning that after these measuring cycles ($|V_W| \leq 7.5$ V) the junctions remain ferroelectric, as shown in Supporting Information S4. However, increasing $V_W$ to 8 V leads to irreversible changes in the junction, which shows a smaller resistance and no evidence of ferroelectric switching (see Supporting Information S5). The presence of two regions (I and II) and similar $R(V_W)$ and $ER(V_W)$ trends have been found in most of junctions in the sample and are insensitive to the duration of the writing pulse ($\tau_W$) and the dwell time ($\tau_D$), within the explored range ($\tau_W = 10$ - 500 µs and $\tau_D = 0.25$ s - 1.0 s) (Supporting Information S6 and S7).

We analyze the shape of the I-V curves to explore the origin of $ER$ in the reversible region I (junction $J_3$). Figure 3c shows the I-V data collected after writing with $V_W = \pm$ 4.5 V, corresponding to the resistance after $\pm$ 4.5 V shown in Figure 2c. I-V data collected after other pulses with $|V_W| = 4.5$ V obtained sequentially are shown in Supporting Information S7. Inspection of data in Figure 3c reveals that: a) I-V curves are sigmoidal and asymmetric as commonly found in trapezoidal tunnel barriers and b) for $V_W^-$ (up triangles) the conductance is larger than for equivalent $V_W^+$ (down triangles), thus reflecting the emergence of *ER*. The I-V curves in Figure 3c were fitted (solid lines) using the Brinkman model[39] to extract the barrier heights ($\Phi_{LSMO}$ and $\Phi_{Pt}$) and the effective barrier width ($t_{eff}$) of a trapezoidal barrier. $\Phi_{LSMO}$ and $\Phi_{Pt}$ refer to the barrier height at the LSMO and Pt electrodes interfaces with the ferroelectric layer, respectively. The mean energy barrier has been calculated as: $\Phi_{mean} = (\Phi_{LSMO} + \Phi_{Pt})/2$. The fitted values are $\Phi_{LSMO} = 1.3$ ($\pm 0.2$) eV, $\Phi_{Pt} = 2.2$ ($\pm 0.2$) eV, $\Phi_{mean} = 1.86$ ($\pm 0.03$) eV and $t_{eff} = 5.9$ ($\pm 0.05$) nm for $V_W = + 4.5$ V. For $V_W = - 4.5$ V, they are $\Phi_{LSMO} = 2.5$ ($\pm 0.1$) eV, $\Phi_{Pt} = 1.4$ ($\pm 0.3$) eV, $\Phi_{mean} = 1.98$ ($\pm 0.07$) eV and $t_{eff} = 5.5$ ($\pm 0.01$) nm. Error bars represent the dispersion of the fitted parameters to I-V data collected after three sequential writing with $V_W = + 4.5$ and $- 4.5$ V. Importantly, repeated cycling of the junction leads to almost identical changes on the barrier properties. I-V data and fitting parameters are shown in Supporting Information S8. These energy barrier values are larger than those found in similar devices using BaTiO$_3$ (BTO) as ferroelectric barrier[41], which is expected due to the lower electron affinity of HfO$_2$ ($\approx 2.0$ eV) compared to BTO ($\approx 3.9$ eV) and fully consistent with the work functions of electrodes.[41,42] According to the sketch of Figure 1a, $V_W > 0$ should impose *P* pointing towards LSMO. This corresponds to a smaller barrier at LSMO ($\Phi_{LSMO} = 1.3$ ($\pm 0.2$) eV) interface and a higher one ($\Phi_{Pt} = 2.2$ ($\pm 0.2$) eV) at Pt side. This produces a trapezoidal barrier as predicted by Zhuravlev et al. [43] and agrees with results obtained in similar BTO ferroelectric junctions.[7,15]

Notice that the height of the energy barriers at LSMO and Pt sides ($\Phi_{LSMO}$, $\Phi_{Pt}$) changes from $\approx$ (1.3 eV, 2.2 eV) to $\approx$ (2.5 eV, 1.4 eV) when reversing the sign of $V_W$. However, the most important here is that the difference $\Delta\Phi = \Phi_{LSMO} - \Phi_{Pt}$ reverses almost symmetrically when changing $V_W$ polarity. Therefore, although $\Delta\Phi$ reverses its sign with *P*, its mean height value $\Phi_{mean}$ remains constant upon *P* reversal. This conclusion, derived from the analysis of the I-V curves, at first sight is at odds with the observation (Figure 2c) that the junction resistance (*R*) is significantly smaller (about 410 %) for $V_W^-$ than for $V_W^+$, in spite of the negligible variation of the mean height barrier energy ($\Phi_{mean}$) with $P$ reversal. The clue to this unexpected observation can be obtained by inspecting the extracted values of $t_{eff}$. As indicated above, for $V_W = + 4.5$ V we obtained $t_{eff} = 5.9$ ($\pm 0.05$) nm but $t_{eff} = 5.5$ ($\pm 0.01$) nm for $V_W^- = - 4.5$ V. This thickness difference ($\Delta t_{eff} \approx 0.4$ nm) accounts for the observation of a low resistance state for $V_W^-$, $P$ pointing away from LSMO. Therefore, upon $V_W$ reversal, the shape of the energy barrier basically reverses symmetrically, but the width shrinks/expands as illustrated in Figure 3d. We conclude that the *ER* observed in Region I can be associated to the modulation of the tunnel barrier properties by polarization reversal. Similar data fitting has been performed in I-V curves collected after $|V_W| = 6$ V (see Supporting Information S8) showing a poorer quality of the fits, which indicates that the tunneling conduction might coexist with other conduction mechanisms as discussed in detail below.

We turn now our focus on the abrupt drop of resistance that signals the transition to Region II. It is worth noticing that this abrupt change of resistance occurs (in Pt/HZO/LSMO//STO) around $\approx 5$ V for both writing voltage pulse polarities ($V_W^+$ and $V_W^-$). This behavior is reminiscent of the forming step observed in $HfO_2$ and other materials in which an electric-field induces the formation of a conducting channel, mostly along grain boundaries.[34] Since the structural mismatch between HZO and the LSMO buffer layer is different for the samples grown on STO and GSO substrates, the formation of monoclinic and orthorhombic phases is different. Indeed, it has been recently reported [40] that the orthorhombic phase is favored with respect to the monoclinic phase in HZO/LSMO samples and substrates with larger lattice parameters than STO ($a = 3.905$Å), as it is the case of GSO ($a = 3.97$ Å). Thus, the HZO film grown on LSMO/GSO contains mainly orthorhombic grains, while the film grown on LSMO/STO is composed of a mixture of orthorhombic and monoclinic grains.

In Figure 4a, 4c and 4e we show HAADF-STEM cross sectional views of HZO/LSMO//STO, HZO/LSMO//GSO and HZO/LSMO//LSAT respectively. Notice that STEM experiments have been conducted on heterostructures containing 9 nm thick HZO layer. A larger field of view in both films can be found in Supporting Information S9. HAADF-STEM images were obtained along the [110] zone axes of the substrates and show a clear contrast between the HZO film and the LSMO electrode. It can be appreciated in Figure 4a that the HZO film grown on LSMO//STO consists of grains with monoclinic and orthorhombic phases, having a lateral size in the 10-15 nm range. The orthorhombic and monoclinic grains are (111) and (001)-textured on the (001) substrate, respectively, which renders non-coherent orthorhombic-monoclinic grain boundaries. They are shown in the higher magnification Z-contrast image of Figure 4b. Notice that the atomic arrangement in these highly mismatched grain boundaries significantly deviates from the structure within the grains which points to a complex atomic reordering at grain boundaries that extends up to several atomic planes. In contrast, the microstructure of the HZO film grown on LSMO//GSO is largely different (Figure 4c). It can be appreciated that the HZO film is also formed by grains, but only (111)-oriented orthorhombic o-HZO grains are present. As expected from the epitaxial growth of a (111) textured film onto (001) cubic substrates, twining is observed, thus the grain boundaries between two adjacent orthorhombic grains, having identical out-of-plane texture, can be in-plane rotated and produce coherent grain boundaries. The in-plane rotation is indicated in the higher magnification Z-contrast image of Figure 4d, where coherent grain boundaries are clearly visualized. Notice that these grain boundaries are thinner and less distorted than in HZO/LSMO//STO. In short, o-HZO and m-HZO crystallites can be identified in some regions of the HZO/LSMO//STO sample (Figure 4a,b) and thus incoherent and highly distorted grain boundaries must exist. In contrast, for HZO/LSMO//GSO (Figure 4c,d), the HZO film displays a homogeneous texture corresponding to o-HZO (111) phase and incoherent m-HZO/o-HZO interfaces are absent. The presence of grain boundaries between monoclinic crystallites is also apparent in the images of HZO/LSMO//LSAT film, as shown in Figure 4e and 4f.

We now analyze the impact of grain boundaries on the junction resistance. We focus our attention on $R(V_W)$ with increasing $V_W$ to identify regions I and II of HZO films on different substrates. Figures 5a and 5c show $R(V_W)$ on the Pt/HZO/LSMO//GSO and Pt/HZO/LSMO//LSAT samples, respectively. The I-V loops for junctions grown on these substrates, indicating its $V_C$, can be seen in Supporting Information S10. We also include (Figure 5b) data for Pt/HZO/LSMO//STO from Figure 2b for comparison. It is clear that the characteristic abrupt decay of resistance occurring at some $V_W$ (around 5 V in Pt/HZO/LSMO//STO) is shifted in Pt/HZO/LSMO//GSO for much larger voltage ($\approx 13$ V). Instead, $R(V_W)$ displays two well defined and distinguishable $R(V_W^+)$ and $R(V_W^-)$ states developing at $V_W \approx 4.5$ V. Their difference remains roughly constant up to $\approx 13$ V, indicating that the energy barrier depends on the sign of the polarizing voltage ($V_W^+$ and $V_W^-$) but not on its magnitude, as expected for a ferroelectric barrier with saturated polarization. Consistently, $ER(V_W)$ displays robust loops ($ER \approx 450$ %) when cycling the junction up to 6 V, as show in Figure 5d. The fingerprint of ion-related $ER$ in Pt/HZO/LSMO//GSO sample is shifted to 13 V (Figure 5d) supports the view that, in this sample, conducting channels along monoclinic-orthorhombic grain boundaries are mostly suppressed. Consistently, the $ER$ data HZO films grown on TbScO$_3$ substrates (Pt/HZO/LSMO//TSO) (see Supporting Information S11) shows only evidence of the polarization related resistive switching and the absence of the characteristic decay of resistance associated to grain boundaries. Figures 5c and 5f show the corresponding data for Pt/HZO/LSMO//LSAT. The overwhelming presence of non-ferroelectric monoclinic phase and residual existence of orthorhombic phase is apparent on the observation of a tinny opening of the $R(V_W)$ loop and a reduced electroresistance (Figure 5f). Both observations are consistent with polarization loops and X-ray data.

Finally, we note that the threshold field at which ER develops ($V_W \approx 4.5$ V) in the Pt/HZO/LSMO//GSO is larger than that observed in Pt/HZO/LSMO//STO (about 2.5 V). If, as argued here, the $ER(V_W)$ in region I is due to polarization reversal then this would imply that the coercive voltage $V_C$ of Pt/HZO/LSMO//GSO is larger. Indeed, this is what we observed (Supporting Information S10).

**Summary**

Here, we have reported exhaustive $ER(V_W)$ and ferroelectric polarization measurements of nanometric (4.6 nm) ferroelectric epitaxial HZO barriers grown on LSMO (bottom electrode) deposited on different substrates (STO, GSO and TSO), which have different lattice mismatch with HZO. We have found that in Pt/HZO/LSMO//STO devices the junction resistance $R(V_W)$ displays two well defined regions (I, II) with dependence on the writing voltage $V_W$. In region I, a substantial $ER$ develops ($\approx 410$ %) giving rise to well defined low/high (ON/OFF) resistance states. Data indicate that the two states are triggered by switching of the ferroelectric polarization of HZO and that the conduction mechanism is tunneling. Therefore, $ER$ develops coinciding with the ferroelectric layer coercive voltage ($V_C$). Interestingly, the polarization reversal affects the barrier energy profile almost symmetrically and mostly modulates the barrier width. This observation suggests that polarization-dependent depletion layers formed at electrode/HZO interfaces have a prominent role in the $ER$. When increasing $V_W$ well above $V_C$, the junctions enter in the so-called region II at some critical voltage where a large and non-fully-reversible $ER$ develops. It is argued that this second voltage threshold is a fingerprint of the presence of a different $ER$ mechanism not related to polarization but to ionic motion. This effect is commonly observed in non-ferroelectric devices based on $HfO_2$ and other oxide thin films and extensively discussed in the literature. We have argued that the presence of specific grain boundaries in HZO is instrumental for the formation of conducting channels that may mask genuine polarization switching $ER$. This hypothesis is conclusively assessed by growing and characterizing the microstructure and comparing *ER* in HZO/LSMO heterostructures on STO and perovskite scandates (GSO and TSO). STEM images with atomic resolution demonstrate that incoherent boundaries (between monoclinic and orthorhombic phases) exist in HZO/LSMO//STO but only coherent grain boundaries are observed in HZO/LSMO//GSO. Consistently, it is found that junctions in HZO/LSMO//GSO remain within region I upon $V_W$ cycling and region II is pushed to larger voltages beyond the explored range ($V_W \approx 8$ V). Consequently, the use of GSO substrates has allowed to obtain larger *ER* values (up to 450 %) displaying a reversible behavior polarization-switchable ferroelectric tunnel devices based on HZO.

**Conclusion**

We have disclosed here the dramatic role that the microstructure, namely the existence of grain boundaries between orthorhombic and monoclinic phases, has on the *ER* of epitaxial HZO tunnel barriers. The message emerging from this work is that polarization-related and ionic-like conductive mechanisms coexist in a tunnel barrier. Both effects give rise to an *ER*, although fully reversible behavior (without forming steps) is only observed in the former. In ferroelectric tunnel devices, the *ER* is commonly rationalized by the change of barrier height upon polarization switching. Interestingly, it has been observed here that, in HZO films, the effective tunneling thickness modulation upon polarization switching rules the tunneling current, indicating the very relevant role of the interface in this system. It has been shown that appropriate substrate selection allows to drastically reduce the grain boundary density, which results in an enhanced and robust polarization-related *ER* (up to 450 %). Probably, reduction of the contact size well below micron size may allow further improvements. These findings should contribute to a faster development and implementation of $HfO_2$-based ferroelectric tunnel devices.

**Experimental Section**

*Sample Growth:* Epitaxial HZO films with nominal thicknesses 4.6 nm were grown on LSMO (22 nm thick) buffered STO (001), (001)-oriented (pseudocubic indexation) GSO and TSO substrates as described elsewhere.[40] Pt electrodes (20 nm thick) were grown ex-situ, at room temperature, by DC-sputtering through shadow mask which allows the deposition of circular top contacts (diameter $\approx$ 20 $\mu$m). Junctions are identified by the code $J_n$. Single crystalline STO, GSO and TSO used as substrates have bulk cell parameters (pseudo-cubic) of 3.905 Å, 3.97 Å and 3.958 Å, respectively. Full structural characterization of HZO films on those substrates has been reported elsewhere.[40]

*Structural Characterization:* X-ray diffraction 2θ-χ recorded using Bruker Bruker-AXS D8 Advance equipped with a Vantec 500 detector (Cu−K$_\alpha$ radiation). The corresponding θ-2θ scan was obtained by integration within the ± 10º range angular.

*Electrical Characterization*: The electrical contact configuration is sketched in Figure 1a. The bottom LSMO layer acts as electrical ground and was contacted through silver paste contact at the edge of the sample. Top Pt electrodes were biased. Polarization measurements were done by integrating the current though time of collected I-V curves at indicated frequency. Electroresistance measurements are done by using a triangular voltage pulse of variable amplitude $V_W(t)$ and duration $\tau_W$ (10 μs − 500 μs) to set the device initial state (writing step). A linear $V_R(t)$ pulse of maximum amplitude 1 V is subsequently used to read the resistance of the junction. Quoted values of junction resistance correspond to $V_R = 0.9$ V. The delay time $\tau_D$ is fixed at 0.5 s. All electrical characterizations were performed with an AixACCT TFAnalyser2000 platform. The $P(V)$ loop in Figure 1d has been obtained by integration of the I-V data in Figure 1c and subtracting the dielectric background (Supporting Information S12).

*Transmission electron microscopy characterization:* Aberration-corrected scanning transmission electron microscopy (STEM) was used for microstructural analysis with atomic resolution. Samples were characterized using a NION UltraStem operated at 200 kV equipped with a NION corrector and a JEOL JEM ARM200cF operated at 200 kV, equipped with a CEOS aberration corrector. The STEM microscopes were operated in high angle annular dark field (HAADF) imaging mode, also referred to as Z-contrast because the brightness associated to each atomic column scales with its atomic number.[44] Specimens for STEM were prepared by conventional methods: grinding, dimpling, and Ar ion milling.

**Acknowledgements**

Financial support from the Spanish Ministry of Economy, Competitiveness and Universities, through the "Severo Ochoa" Programme for Centres of Excellence in R&D (SEV-2015-0496) and the MAT2017-85232-R (AEI/FEDER, EU), and MAT2015-73839-JIN projects, and from Generalitat de Catalunya (2017 SGR 1377) is acknowledged. MC acknowledges fellowship from "la Caixa Foundation" (ID 100010434; LCF/BQ/IN17/11620051). IF and JG acknowledge Ramón y Cajal contracts RYC-2017-22531 and RYC-2012-11709, respectively. SE acknowledges the Spanish Ministry of Economy, Competitiveness and Universities for his PhD contract (SEV-2015-0496-16-3). JL and XL are financially supported by China Scholarship Council (CSC) with No. 201506080019 and 201806100207. MC work has been done as a part of her Ph.D. program in Physics at Universitat Autònoma de Barcelona. SE and JL work has been done as a part of their Ph.D. program in Materials Science at Universitat Autònoma de Barcelona. Authors acknowledge the ICTS-CNME for offering access to their instruments and expertise. The electron microscopy performed at ORNL was supported by the Materials Sciences and Engineering Division of Basic Energy Sciences of the Office of Science of the U.S. Department of Energy.

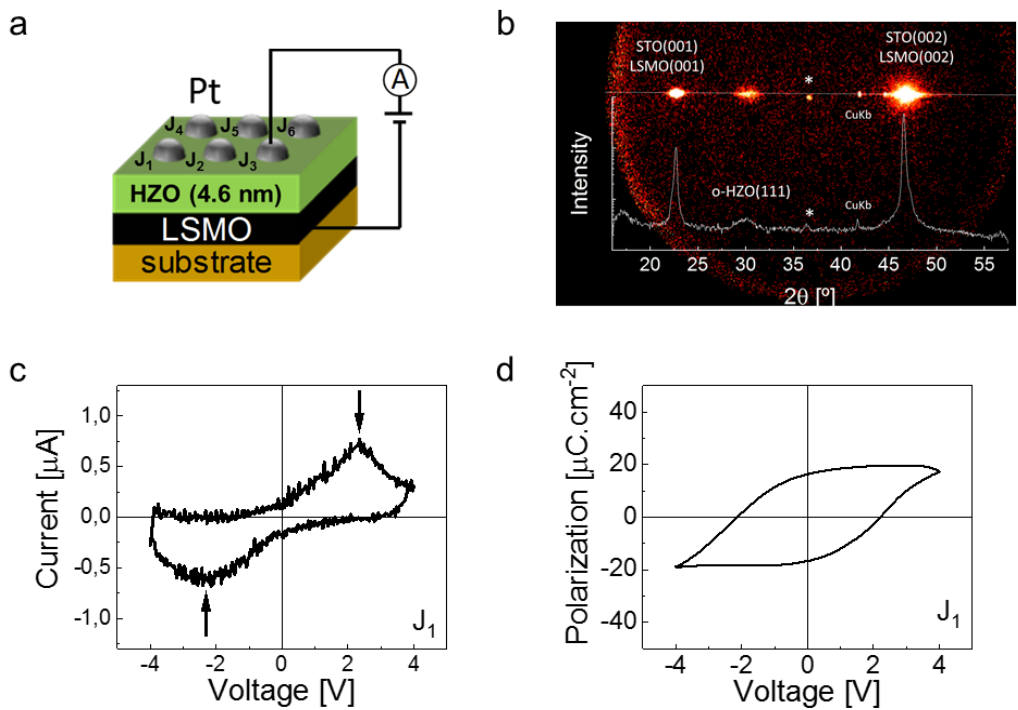

**Figure 1.** (a) Sketch of the sample heterostructures and contact arrangement for electrical measurements. (b) X-ray diffraction θ-χ frame of the HZO/LSMO//STO sample and the corresponding θ-2θ scan. The intense (00l) reflections of substrate and LSMO can be appreciated as well as the (111) reflection of the ferroelectric orthorhombic HZO. The peak labeled with * is an artifact from the detector. (c) Current-voltage loops ($V_{max} = 4$ V) of junction $J_1$ recorded at 1 kHz. (d) The corresponding polarization $P(V)$ loop obtained.

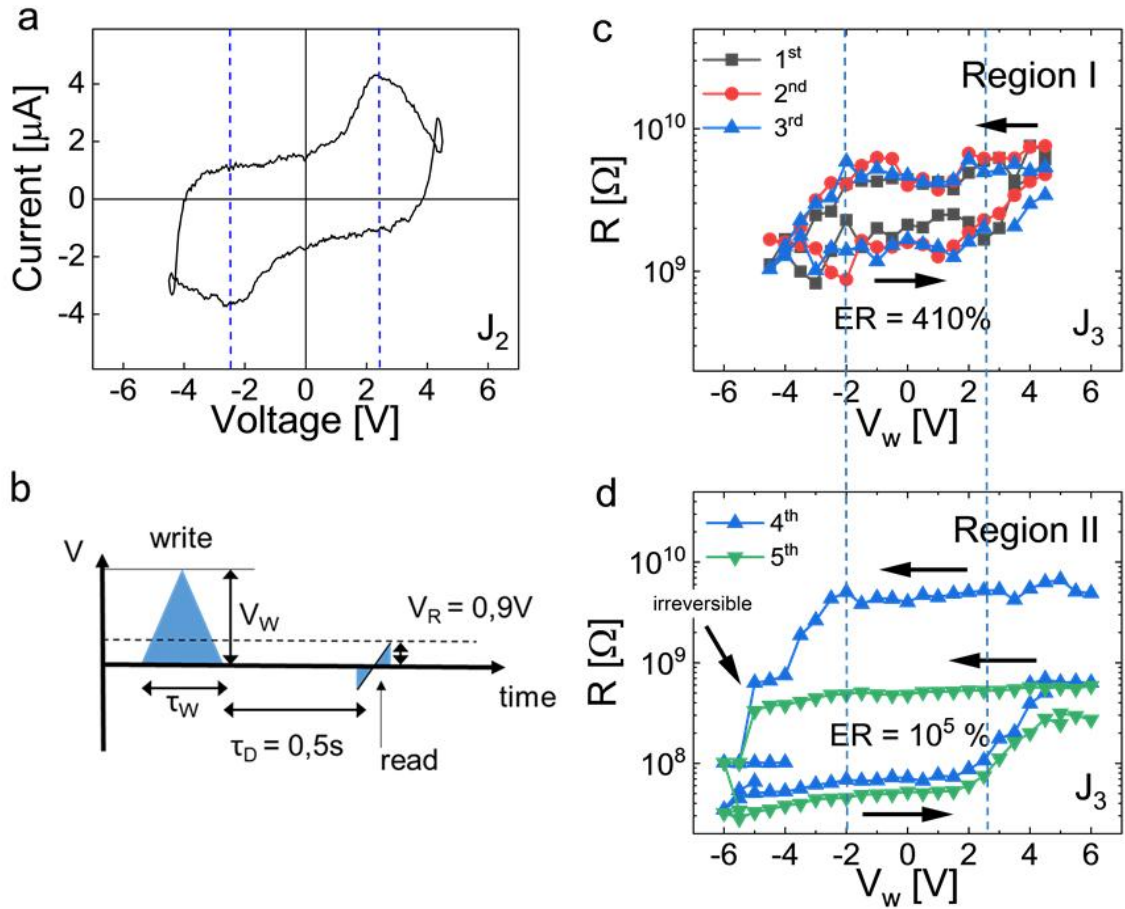

**Figure 2.** (a) I-V loops of Pt/HZO/LSMO//STO measured at 5 kHz before performing $R(V_W)$ measurement cycles up to $V_W = 4.5$ V. (b) Sketch of the writing/reading measuring protocol. The amplitude of the writing pulse ($V_W$), its duration ($\tau_W$) and dwell time ($\tau_D$) before the measuring pulse of amplitude $V_R$ are indicated. (c) Dependence of the junction resistance (R) on the writing voltage $V_W$ ($|V_W^{+,-}| < 4.5$ V); three consecutive cycles are shown to illustrate reversibility. (d) $R(V_W)$ cycles for $|V_W^{+,-}| < 6$ V; two consecutive cycles are shown illustrating irreversibility. Dashed vertical lines indicate the coercive voltage ($V_C$) of the HZO film extracted from (a).

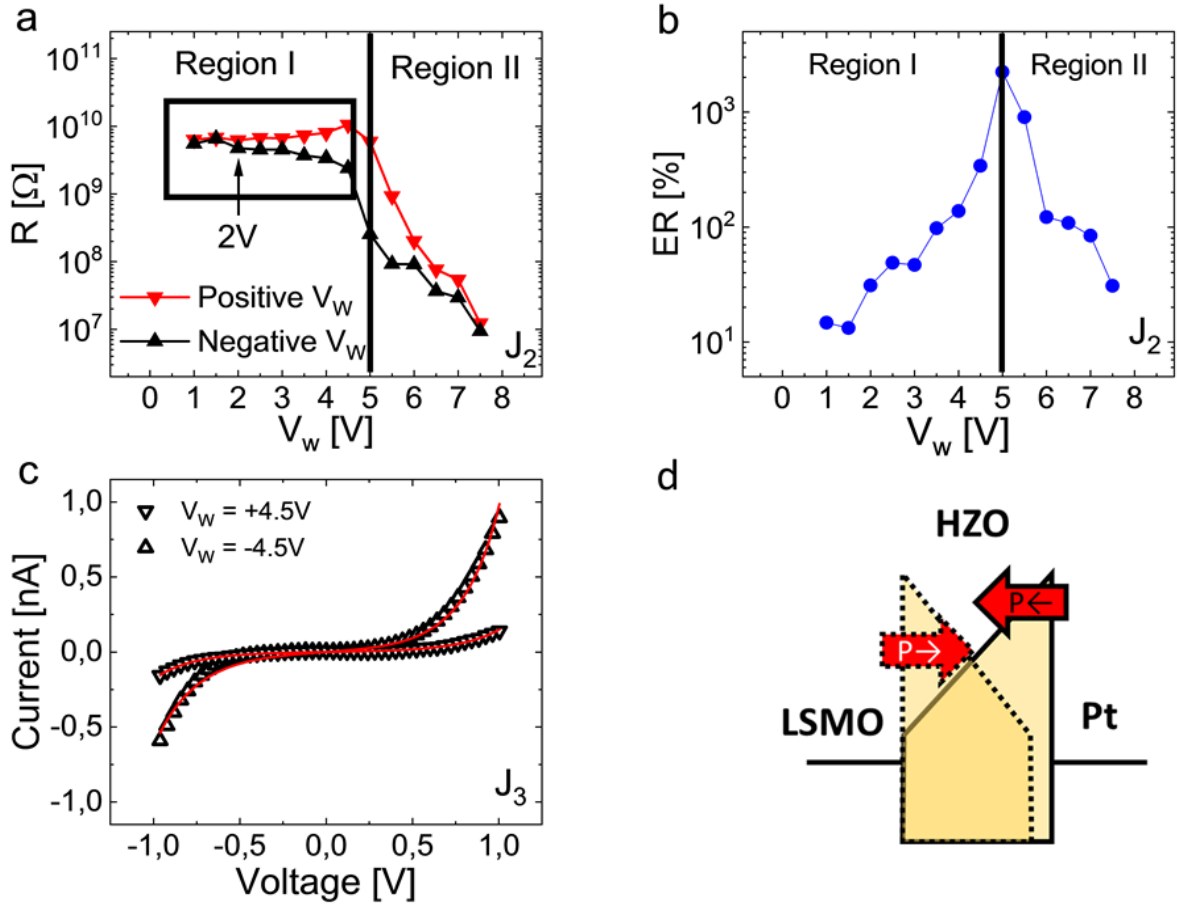

**Figure 3**. (a) Resistance (at 0.9 V) of the device (J$_2$) in Pt/HZO/LSMO//STO sample after positive ($V_W > 0$) and negative ($V_W < 0$) writing pulses. (b) Electroresistance calculated from data shown in (a). (c) I-V curves measured (symbols) in Pt/HZO/LSMO//STO (junction J$_3$) after writing with $V_W = + 4.5$ and $- 4.5$ V, as indicated. The solid red lines are the results of the fit as described in the text. (d) Sketch of the energy barrier profile of the device, upon $P$ reversal, based on the $\Phi_{LSMO}$ and $\Phi_{Pt}$ energy barrier heights and thickness parameters derived from the fits of $I$(V) curves in Region I. For convenience, the effective thickness decrease upon polarization reversal has been sketched in the Pt/HZO side, although it can be also probable to be produced in the LSMO/HZO side or both.

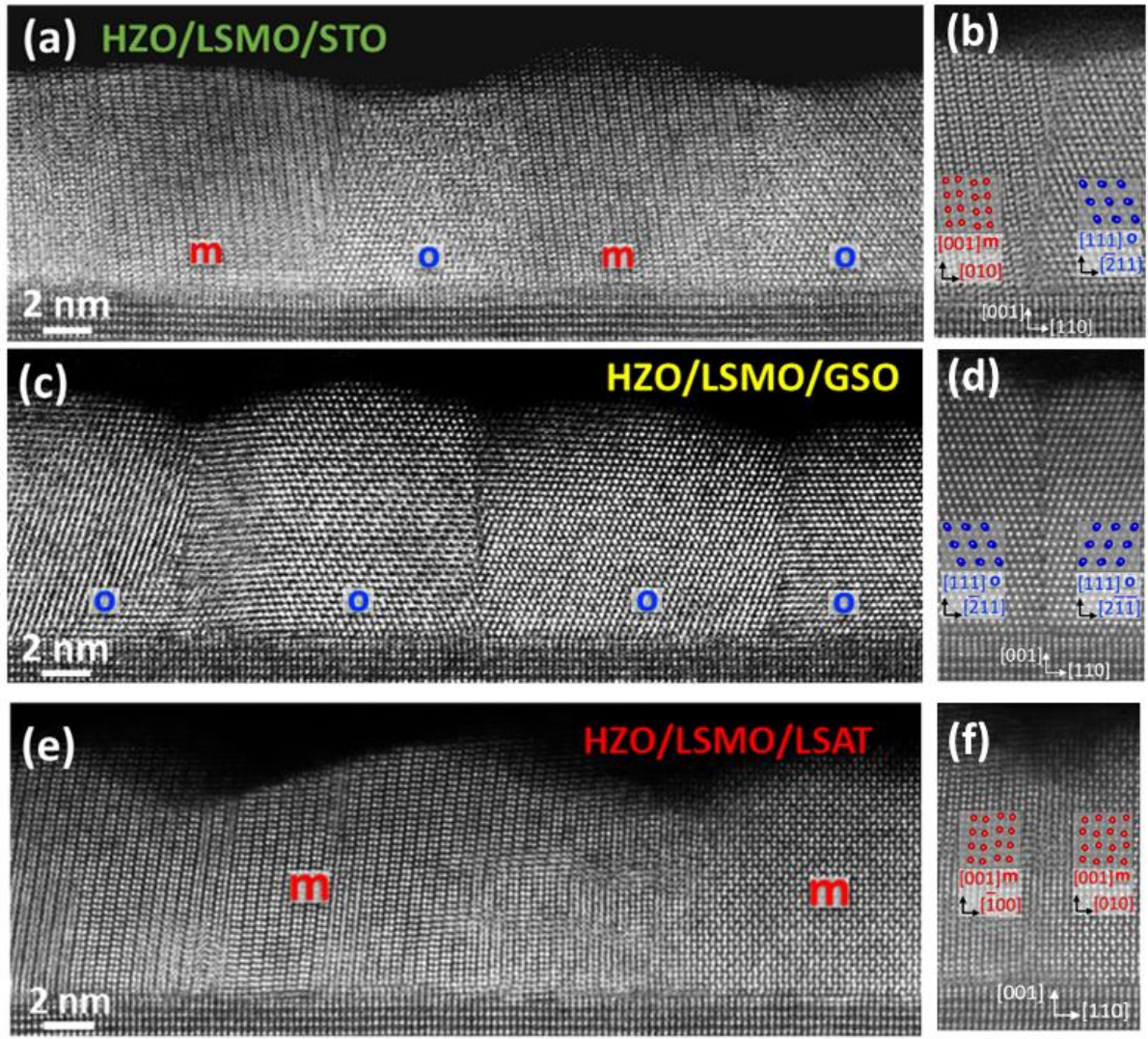

**Figure 4.** (a, c, e) Cross-sectional atomic-resolution Z-contrast images of 9nm-thick HZO films grown onto HZO/LSMO//STO, HZO/LSMO//GSO and HZO/LSMO//LSAT heterostructures, respectively. The images were acquired along the [110] zone axes of the substrates. Monoclinic grains are signaled in red and orthorhombic grains in blue. (b, d, f) Higher magnification Z-contrast images of the grain boundary. Red and blue circles depict the projected monoclinic (space group P2$_1$/c) and orthorhombic (space group Pca2$_1$) structures, respectively. The epitaxial relationship for the monoclinic phase is [010]m-HZO(001)//[110]LSMO(001), [001]m-HZO(001)//[001]LSMO(001) and [100]m-HZO(001)//[1-10]LSMO(001), while for the orthorhombic phase is [-211]o-HZO(111)//[110]LSMO(001) and [111]o-HZO(001)//[001]LSMO(001), where all the indices

refer to the cubic or pseudocubic unit cells. LSMO and the substrates have a cube-on-cube epitaxial relationship.

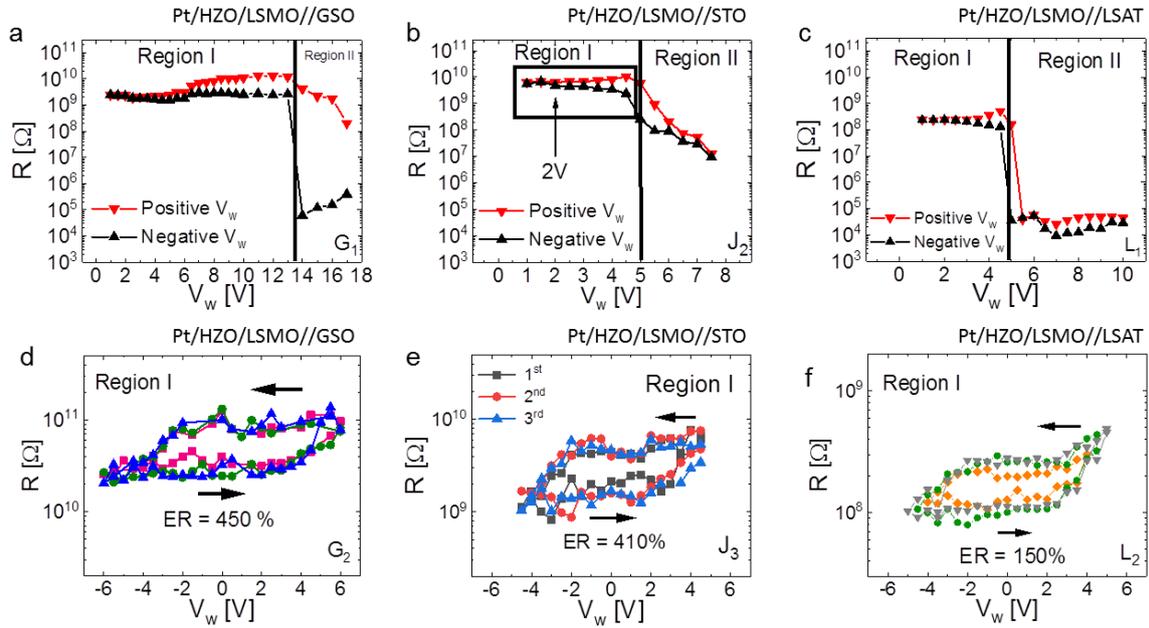

**Figure 5.** (a, b, c) Dependence of the resistance on the writing voltage $V_W$ [$R(V_W)$] in Pt/HZO/LSMO junctions grown on different substrates as indicated. The Pt/HZO/LSMO//STO data (taken from Figure 2a) are included for comparison. (d, e, f) Electroresistance $R(V_W)$ loops measured in junctions with structure indicated.

Supporting Information

# Unraveling ferroelectric polarization and ionic contributions to electroresistance in epitaxial $Hf_{0.5}Zr_{0.5}O_2$ tunnel junctions

Milena Cervo Sulzbach[1], Saúl Estandía[1], Xiao Long[1], Jike Lyu[1], Nico Dix[1], Jaume Gàzquez[1], Matthew F. Chisholm[2], Florencio Sánchez[1], Ignasi Fina[1]*, Josep Fontcuberta[1]*

[1] Institut de Ciència de Materials de Barcelona (ICMAB-CSIC), Campus UAB, Bellaterra, Catalonia 08193, Spain

[2] Center for Nanophase Materials Sciences, Oak Ridge National Laboratory, Tennessee 37831-6064, USA

**Supporting Information S1.**

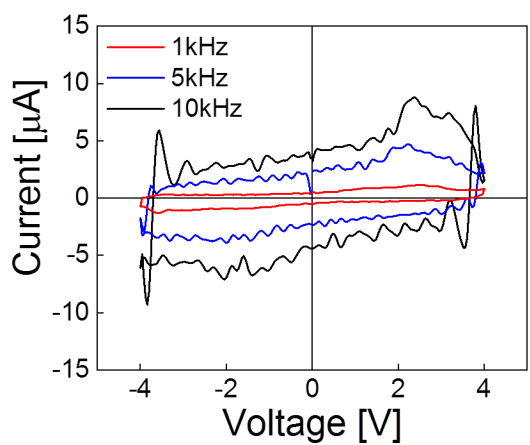

Figure S1. $I(V)$ loops of sample Pt/HZO/LSMO//STO recorded at different frequencies. The emergence of spurious peaks at the maximum voltage in the $V(t)$ excursion when recorded at high frequency is illustrated.

**Supporting Information S2.**

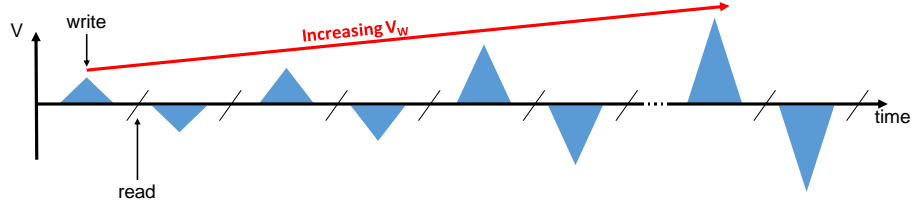

Figure S2. Voltage excursion used to write the junctions with $V_\mathrm{w}$ pulses of increasing amplitude.

**Supporting Information S3.**

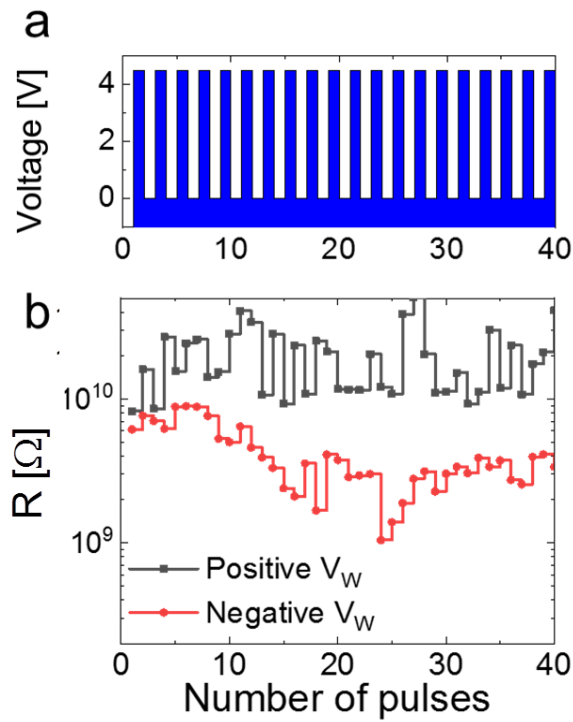

Figure S3. (a) Consecutive equal polarity pulses sequence of $V_\mathrm{w} = 4.5$ V and $\tau_\mathrm{w} = 100$ μs in Junction J$_4$. (b) Resistance measured after applying indicated number of pulses of the same polarity. 40 consecutive prepoling pulses of polarity opposite to the one indicated for each set of data was applied prior to the experiment. The data show that certain number of pulses are needed to well-set the state of lower resistance, although important stochastic response is also observed.

**Supporting Information S4.**

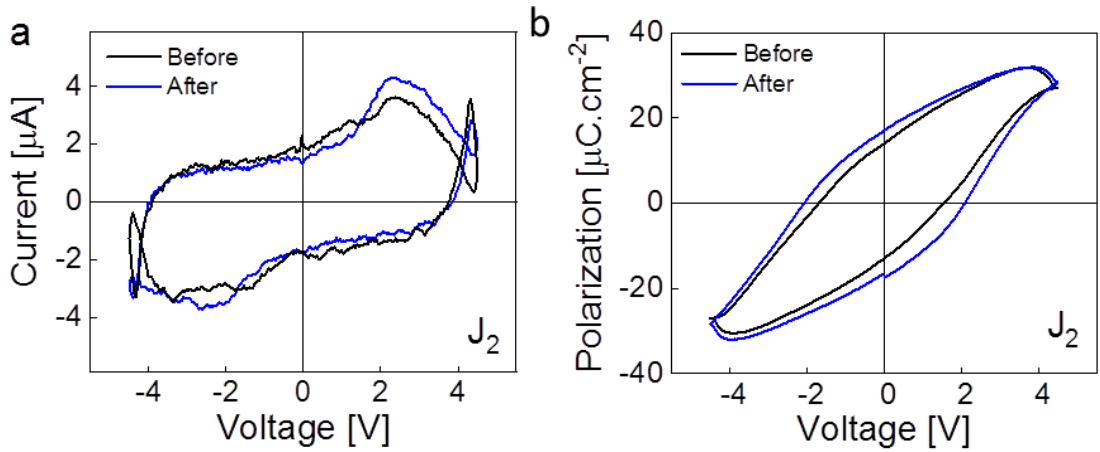

Figure S4. (a) I-V curves and (b) $P(V)$ loops recorded after a writing step of $V_w$ = 7.5 V on junction $J_2$ of Pt/HZO/LSMO//STO sample shown in Figure 3a,b (main text).

**Supporting Information S5.**

Current-voltage loops after measuring $ER$ cycles ($|V_W| \leq 7.5$ V) in which the junction remains ferroelectric. However, increasing $V_W$ to $\approx 8$ V leads to irreversible changes which display a much smaller resistance and no evidence of ferroelectric switching.

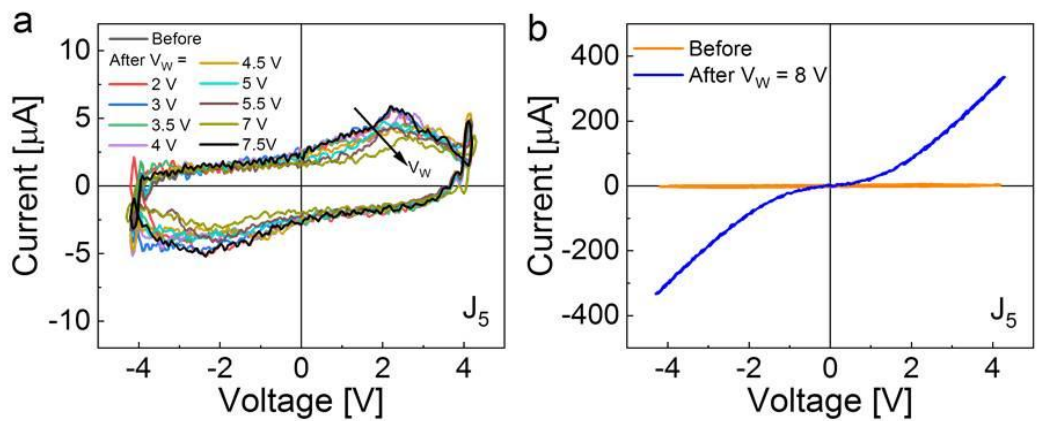

Figure S5. (a) I-V loops collected (junction $J_5$) at 5 kHz; this frequency is selected to coincide with $1/\tau_W$ used in Figure 2c,d and 3a,b. (b) Comparison of the I-V loop, prior (orange data) and after (blue data) a $V_W$ = 8 V writing voltage. Data is collected in junction $J_5$ of Pt/HZO/LSMO//STO sample.

**Supporting Information S6.**

The existence of regions I and II in $R(V_W)$ and $ER(V_W)$ is found in most junctions of Pt/HZO/LSMO//STO sample. Here we show data collected on different junctions on the same sample.

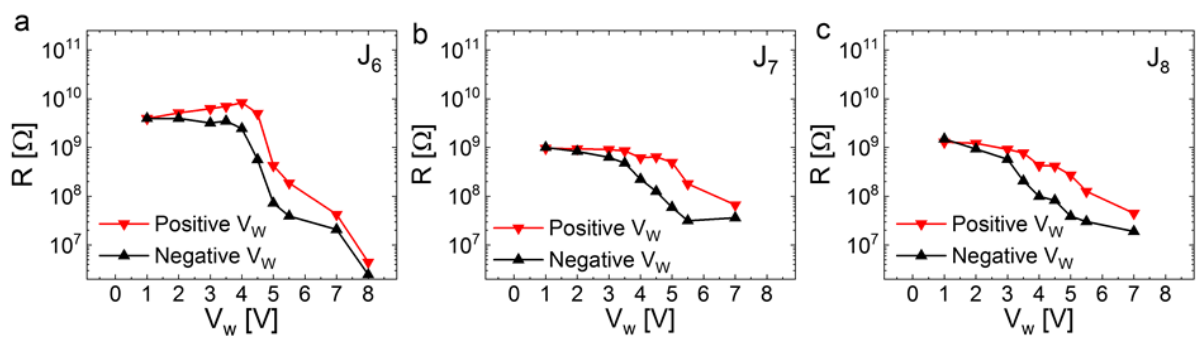

Figure S6. (a,b,c) Electroresistance measured at different junctions ($J_6$, $J_7$, $J_8$) on the same Pt/HZO/LSMO//STO sample. In (a) clear I and II regions can be identified. In (b) and (c) we show examples on junctions where region I is not evident by only the transition to region II. One can notice in (b) and (c) that the voltage in which the transition to region II happens is definitely smaller than in (a) and also the overall resistance of these junctions is smaller. This is consistent with the existence of non-tunnel conducting channels under these electrodes.

**Supporting Information S7.**

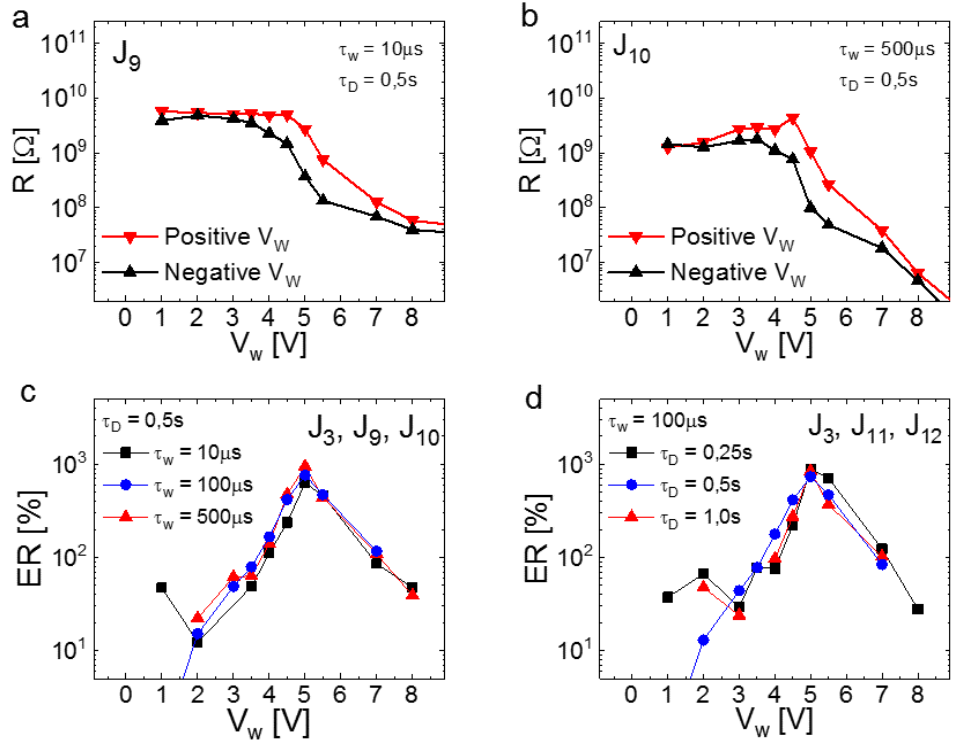

Figure S7. (a,b) Raw data of junction resistance measured after writing pulses of $V_W$ amplitude as indicated using writing time $\tau_W = 10$ µs (junction $J_9$) and 500 µs (junction $J_{10}$), respectively and $\tau_D = 0.5$ s. (c) Electroresistance measured at $\tau_W = 10$ µs, 100 µs, and 500 µs (junctions $J_9$, $J_3$, $J_{10}$, respectively). (d) Electroresistance measured at $\tau_W = 100$ µs and $\tau_D = 0.25$ s, 0.5 s and 1.0 s (junctions $J_{11}$, $J_3$, $J_{12}$, respectively). Data indicate that, within the explored range, the transition from region I to region II is not sensitive to $\tau_W$ and $\tau_D$.

**Supporting Information S8.**

It can be inferred that the fittings to the data collected after $V_W = \pm 4.5$ V are in good agreement (Figure S8a,b). Instead, the fitting quality of data collected after $V_W = \pm 6$ V is worse, see the deviation signaled by an arrow in Figure S8c. In any case, the $\Phi_{LSMO}$, $\Phi_{Pt}$ and $t_{eff}$ parameters (summarized in Table S1) are reasonable for a ferroelectric tunneling device.

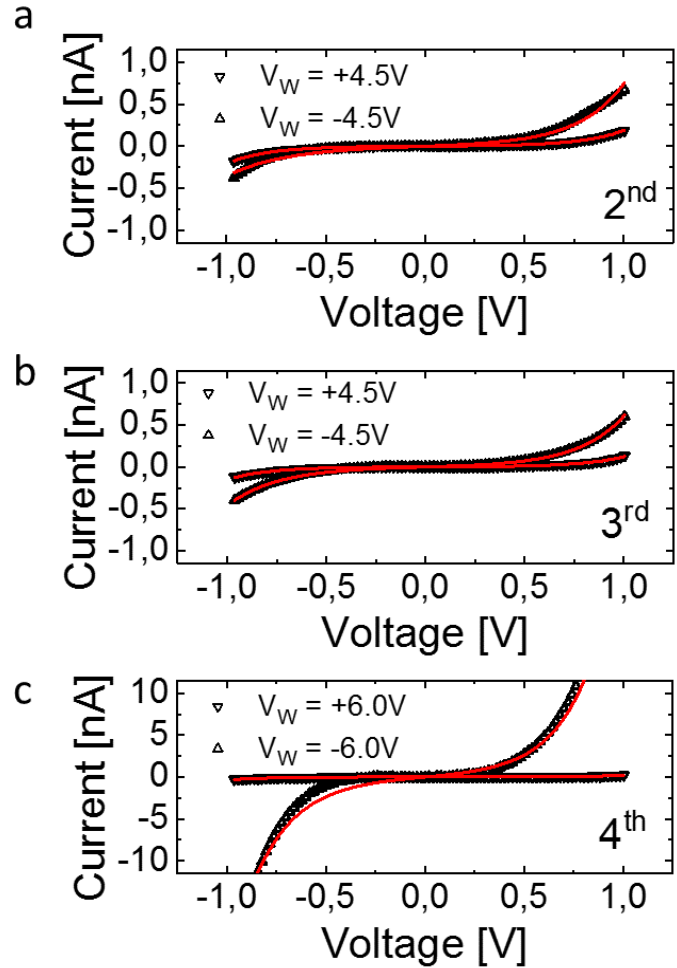

Figure S8. (a,b,c) I-V curves obtained after repeated $V_W$ cycling of sample in the order labeled in the panel and the $V_w$ indicated. Solid lines indicate the results of the fit using the Brinkman model [39] as in [7].

The fittings are obtained by minimizing the $\chi^2$ [$\chi^2 = \Sigma$(experimental-fitted)$^2$] value accounting for the difference between the experimental values and the fitted ones using the equation:

$$J \cong C \frac{exp\left\{\alpha(V)\left[\left(\varPhi_2 - \frac{eV}{2}\right)^{\frac{3}{2}} - \left(\varPhi_1 + \frac{eV}{2}\right)^{\frac{3}{2}}\right]\right\}}{\alpha^2(V)\left[\left(\varPhi_2 - \frac{eV}{2}\right)^{\frac{1}{2}} - \left(\varPhi_1 + \frac{eV}{2}\right)^{\frac{1}{2}}\right]^2} \times sinh\left\{\frac{3}{2}\alpha(V)\left[\left(\varPhi_2 - \frac{eV}{2}\right)^{\frac{1}{2}} - \left(\varPhi_1 + \frac{eV}{2}\right)^{\frac{1}{2}}\right]\frac{eV}{2}\right\} \quad (1)$$

where

$$C = -\frac{4em^*m_e}{9\pi^2\hbar^3}$$

$$\alpha(V) \equiv \frac{\left[4d(2m^*m_e)^{\frac{1}{2}}\right]}{[3\hbar(\varPhi_1 + eV - \varPhi_2)]}$$

and $m_e$ corresponds to the electron mass, $m^*$ to its effective mass (fixed to 0.1) and $e$ to its charge.

| $V_{\mathrm{w}}$ [V] | $t_{\mathrm{eff}}$ [nm] | $\varPhi_{\mathrm{LSMO}}$ [eV] | $\varPhi_{\mathrm{Pt}}$ [eV] | $\varPhi$ [eV] | $R$ [GΩ] |
|---|---|---|---|---|---|
| 4.5 | 5.7 | 1.8 | 2.3 | 2.0 | 7.2 |
| -4.5 | 5.6 | 2.4 | 1.5 | 1.9 | 1.1 |
| 4.5 | 6.0 | 1.6 | 2.1 | 1.9 | 7.6 |
| -4.5 | 5.4 | 2.4 | 1.7 | 2.1 | 1.7 |
| 4.5 | 6.0 | 1.7 | 2.0 | 1.8 | 5.4 |
| -4.5 | 5.6 | 2.7 | 1.2 | 1.9 | 1.0 |
| 6.0 | 5.9 | 1.3 | 2.5 | 1.9 | 4.9 |
| -6.0 | 5.5 | 1.8 | 1.4 | 1.6 | 0.0 |

Table S1. Collected parameters ($\varPhi_{\mathrm{LSMO}}$, $\varPhi_{\mathrm{Pt}}$ and $t_{\mathrm{eff}}$) extracted from the fit of the I-V curves of Figure S8 and data in Figure 3c (main text) to the Brinkman equation (Equation 1) recorded after different writing voltages $V_{\mathrm{w}}$.

**Supporting Information S9.**

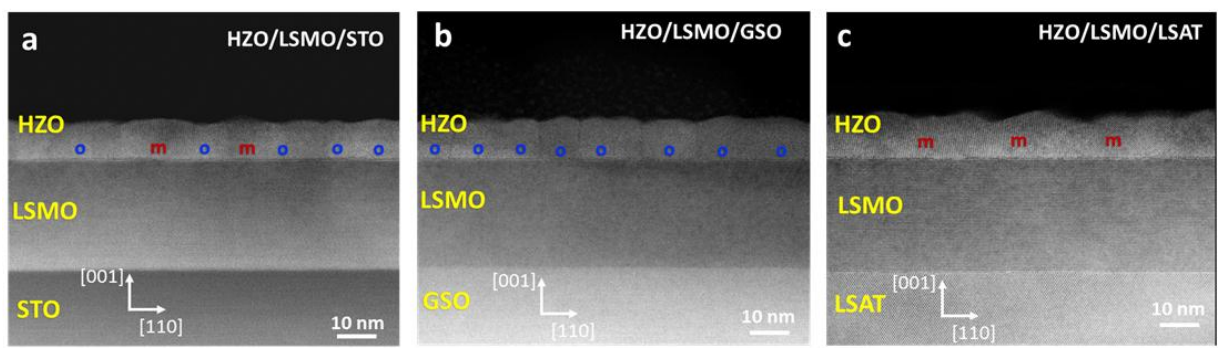

Figure S9. (a,b) Large field of view Z-contrast images of (a) HZO/LSMO//STO, (b) HZO/LSMO//GSO and (c) HZO/LSMO//LSAT. HZO thin film, LSMO buffer layer and STO/GSO/LSAT substrates are visible and produce different contrasts due to the different atomic number Z of elements present in each layer.

**Supporting Information S10.**

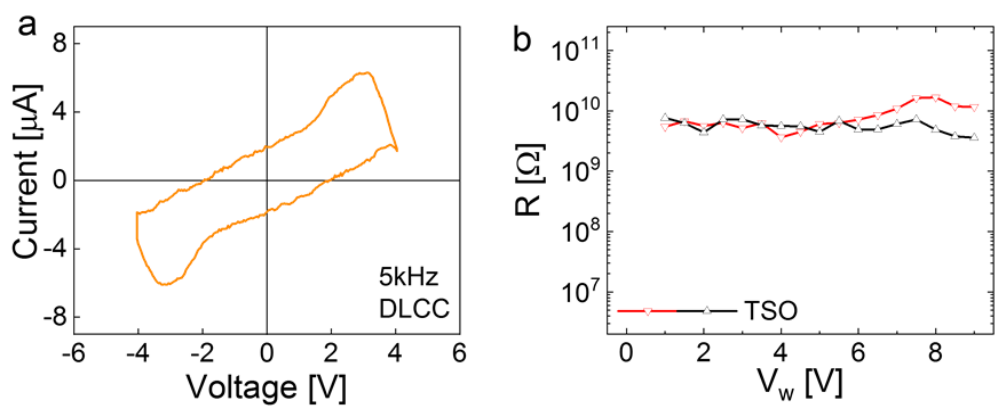

Figure S10. (a) I-V curve recorded in Pt/HZO/LSMO//TSO sample. (b) Dependence of the resistance of one junction in Pt/HZO/LSMO//TSO with the writing voltage: $V_W^+$ (down triangles, red) and $V_W^-$ (up triangles, black).

**Supporting Information S11.**

I-V loops measured in junctions grown on different substrates with configuration presented in Figure 1a (in the main text) are shown in Figure S11. The measurement in Pt/HZO/LSMO//STO and Pt/HZO/LSMO//GSO junctions were performed with an automatic compensation mode (DLCC). This mode assumes that the leakage is independent on the measurement frequency and it is subtracted from the loop. This mode successfully subtracted the leakage contribution from junctions on STO and GSO substrates, however, for LSAT, it distorts the loop. Therefore, Figure S11 shows the loop of the HZO film on LSAT recorded without compensation. A small ferroelectric switching peak can be observed at coercive voltages $V_C^+ \approx +2$ V and $V_C^- \approx -1.8$ V; the ferroelectric response is smaller than in junctions on STO and GSO as expected, since the fraction of orthorhombic (FE) grains should be definitely smaller in HZO films on LSAT than on GSO and even STO. Indeed, in the LSAT only monoclinic (no-FE) grains are visible in the STEM images of the explored region (Figure S9). More details can be found in Ref. 40 in the main text.

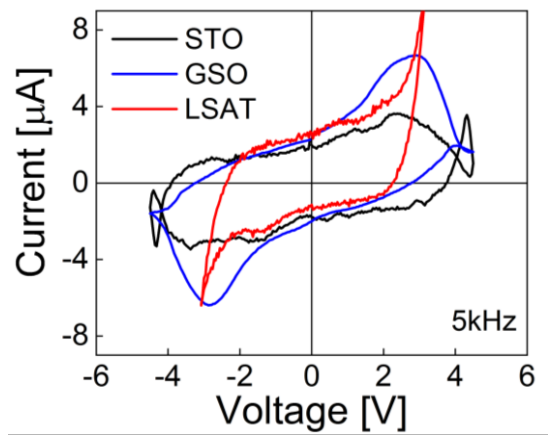

Figure S11. I-V loops measured in Pt/HZO/LSMO//STO, Pt/HZO/LSMO//GSO and Pt/HZO/LSMO//LSAT films. Films on STO and GSO were recorded using the DLCC compensation protocol. Even though the junction on LSAT shows a large leakage, the small ferroelectric switching peak can be seen around 2 V. Coercivity of HZO films on GSO is larger than in HZO on STO.

**Supporting Information S12.**

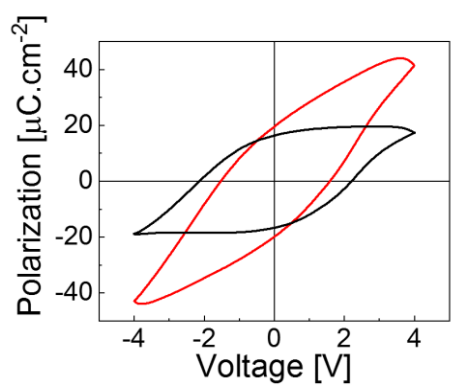

Figure S12. The black line corresponds to the polarization versus electric field loop measured at 1 kHz and show in Figure 1 (main text) after compensating the residual leakage and dielectric contribution. The red line corresponds to the loop obtained at 1 kHz using the dynamic leakage current compensation technique (DLCC) before compensation. The final compensated loop corresponds to the loop obtained after removing the electric susceptibility contribution by linear subtraction of the constant slope (corresponding to $\varepsilon_r = 31$). Data was collected on junction $J_1$ of Pt/HZO/LSMO//STO sample. Residual leakage exponential contribution (after DLCC) was first fitted and after removed from the measured current.[45]